\documentclass[aps,prd,twocolumn]{revtex4-1}
\usepackage[utf8]{inputenc}
\usepackage{dsfont}
\usepackage{amsfonts}
\usepackage{amsmath}
\allowdisplaybreaks[4]        
\usepackage{amssymb}
\usepackage{euscript}     
\usepackage{braket}
\usepackage{starfont}
\usepackage{color,soul}         
\usepackage{tensor}        
\usepackage{amsthm}
\usepackage{graphicx}
\usepackage{slashed}
\usepackage{leftidx}
\usepackage{subfigure}
\usepackage{bbm}
\definecolor{outerspace}{rgb}{0.25, 0.29, 0.3}
\definecolor{scarlet}{rgb}{1.0, 0.13, 0.0}
\usepackage[header,title,page,titletoc]{appendix}  
\definecolor{princetonorange}{rgb}{1.0, 0.56, 0.0}
\definecolor{WildStrawberry}{rgb}{1.0, 0.26, 0.64}
\definecolor{rossocorsa}{rgb}{0.83, 0.0, 0.0}
\definecolor{navyblue}{rgb}{0.0, 0.0, 0.5}
\usepackage{float}
\usepackage[paper=letterpaper,margin=1in]{geometry}
\usepackage{mathtools}

\DeclareMathAlphabet{\pazocal}{OMS}{zplm}{m}{n}
\newcommand{\req}[1]{(\ref{#1})} 

\newcommand{\bea}{\begin{eqnarray}}

\newcommand{\eea}{\end{eqnarray}}
\newcommand{\ba}{\begin{eqnarray}}
\newcommand{\ea}{\end{eqnarray}}

\newcommand{\be}{\begin{equation}}
\newcommand{\ee}{\end{equation} }
\newcommand{\beqa}{\begin{eqnarray}}
\newcommand{\eeqa}{\end{eqnarray}}
\newcommand{\beqar}{\begin{eqnarray*}}
\newcommand{\eeqar}{\end{eqnarray*}}

\renewcommand{\req}[1]{(\ref{#1})}

\newcommand{\eg}{{\it e.g.,}\ }
\newcommand{\ie}{{\it i.e.,}\ }

\makeatletter
\newcommand{\dal}{\mathop{\mathpalette\dal@\relax}}
\newcommand{\dal@}[2]{%
  \begingroup
  \sbox\z@{$\m@th#1\square$}%
  \dimen0=\fontdimen8
    \ifx#1\displaystyle\textfont\else
    \ifx#1\textstyle\textfont\else
    \ifx#1\scriptstyle\scriptfont\else
    \scriptscriptfont\fi\fi\fi3
  \makebox[\wd\z@]{%
    \hbox to \ht\z@{%
      \vrule width \dimen0
      \kern-\dimen0
      \vbox to \ht\z@{
        \hrule height \dimen0 width \ht\z@
        \vss
        \hrule height 2\dimen0
      }%
      \kern-2.5\dimen0
      \vrule width 2.5\dimen0
    }%
  }%
  \endgroup
}
\makeatother

\usepackage{hyperref}
\hypersetup{
    colorlinks,
    citecolor=navyblue,
    filecolor=navyblue,
    linkcolor=navyblue,
    urlcolor=navyblue
}

\begin{document}

\title{Regular Black Holes From Pure Gravity}
\author{Pablo Bueno}
\email{pablobueno@ub.edu}
\affiliation{Departament de F\'isica Qu\`antica i Astrof\'isica, Institut de Ci\`encies del Cosmos\\
 Universitat de Barcelona, Mart\'i i Franqu\`es 1, E-08028 Barcelona, Spain }

\author{Pablo A. Cano}
\email{pablo.cano@icc.ub.edu}
\affiliation{Departament de F\'isica Qu\`antica i Astrof\'isica, Institut de Ci\`encies del Cosmos\\
 Universitat de Barcelona, Mart\'i i Franqu\`es 1, E-08028 Barcelona, Spain }

\author{Robie A. Hennigar}
\email{robie.hennigar@icc.ub.edu}
\affiliation{Departament de F\'isica Qu\`antica i Astrof\'isica, Institut de Ci\`encies del Cosmos\\
 Universitat de Barcelona, Mart\'i i Franqu\`es 1, E-08028 Barcelona, Spain }


\begin{abstract}

We show via an explicit construction how an infinite tower of higher-curvature corrections generically leads to a resolution of the Schwarzschild singularity in any spacetime dimension $D \ge 5$. 
The theories we consider have two key properties that ensure the results are general and robust: (1)  they provide a basis for (vacuum) gravitational effective field theory in five and higher-dimensions, (2) for each value of the mass, they have a unique static spherically symmetric solution. 
 We present several exact solutions of the theories that include the Hayward black hole and metrics similar to the Bardeen and Dymnikova ones. Unlike previous constructions, these regular black holes arise as vacuum solutions, as we include no matter fields whatsoever in our analysis. We show how the black hole thermodynamics can be studied in a completely universal and unambiguous way for all solutions.


\end{abstract}
\maketitle

{\bf Introduction}.  The existence of spacetime singularities is a generic prediction of general relativity (GR)~\cite{Hawking:1973uf, Senovilla:1998oua}. Identifying the mechanism by which they get resolved in the context of a quantum theory of gravity is one of the fundamental problems of theoretical physics. 

Working under the assumption that the metric description of spacetime survives beyond the GR regime, numerous models of regular --- \ie singularity-free --- black holes have been presented in the literature. These proposals typically rely on introducing \emph{ad hoc} matter, namely, they postulate the form of a regular black hole metric and then determine what stress tensor would be required to sustain it \cite{Sakharov:1966aja,1968qtr..conf...87B,1981NCimL......161G,Dymnikova:1992ux,Borde:1994ai,Hayward:2005gi,Lemos:2011dq,Bambi:2013ufa,Simpson:2018tsi,Rodrigues:2018bdc}. Energy conditions are often violated in such constructions, although this is not a necessary condition \cite{Mars:1996khm,Borde:1996df}. Many of these regular black holes can be embedded in models of  GR coupled to non-linear electrodynamics \cite{Ayon-Beato:1998hmi,Bronnikov:2000vy,Ayon-Beato:2000mjt,Bronnikov:2000yz,Ayon-Beato:2004ywd,Dymnikova:2004zc,Berej:2006cc,Balart:2014jia,Fan:2016rih,Bronnikov:2017sgg,Junior:2023ixh}. These scenarios require highly peculiar Lagrangians  which, in addition, involve tuning some of the theory parameters in terms of integration constants of the solutions.  Alternative mechanisms and scenarios include \cite{Frolov:1989pf,Barrabes:1995nk,Nicolini:2005vd,Ziprick:2010vb,Olmo:2012nx,Balakin:2015gpq,Bazeia:2015uia,Chamseddine:2016ktu,Bambi:2016xme,Sajadi:2017glu, Bejarano:2017fgz,Colleaux:2017ibe,Cano:2018aod,Aros:2019quj,Colleaux:2019ckh,Cano:2020qhy,Cano:2020ezi,Guerrero:2020uhn,Bueno:2021krl,Sajadi:2021ilt,Brandenberger:2021jqs,Olmo:2022cui,Biasi:2022ktq,Junior:2024xmm}. 
For more detailed reviews of previous attempts at constructing regular black holes, see \eg \cite{Ansoldi:2008jw,Lemos:2011dq}.

In this paper we present new families of regular black holes in $D \geq 5$ spacetime dimensions. These describe modifications of the usual Schwarzschild black hole and are solutions of a well-known class of metric theories of gravity  --- known as Quasi-topological gravities \cite{Oliva:2010eb,Quasi,Dehghani:2011vu,Ahmed:2017jod,Cisterna:2017umf} --- which correct the Einstein-Hilbert action with higher-curvature terms. These theories are simply a subset of all possible higher-curvature Lagrangians particularly amenable for analytic computations. However, as we argue, they provide a broad enough set of theories that allows one to hope that the conclusions drawn from them could be general. 
We find that, with a finite number of corrections, these theories generically reduce the degree of divergence of a black hole singularity.
By including an infinite number of them, we find that the resulting black hole solutions become singularity-free as long as the relative gravitational couplings of increasingly higher-order densities satisfy certain mild (and qualitative) conditions. Remarkably, our regular black holes are unique --- they are the only static and spherically symmetric solutions of the corresponding theories. Moreover, there is good evidence to suggest a basis of Quasi-topological theories can be chosen such that there is always a Birkhoff theorem, meaning our solutions are also the unique spherically symmetric solutions of those theories.

We present a few analytic examples, which include a general-dimension version of the Hayward black hole  \cite{Hayward:2005gi}  as well as others reminiscent of the Bardeen \cite{1968qtr..conf...87B} and Dymnikova ones \cite{Dymnikova:1992ux}. We comment on the thermodynamic properties of the solutions, which can be analyzed in full generality. We show that the solutions satisfy the first-law of thermodynamics in a completely unambiguous  fashion --- as opposed to previous studies of regular black holes. 

To the best of our knowledge, our solutions represent the first instances of regular black hole solutions to purely gravitational theories in which no special fine-tuning  or constraints between the relevant parameters is required.  The Schwarzschild black hole singularity is generically resolved in a fully controlled setup by the effect of an infinite tower of higher-curvature densities, with no tricks involved.

{\bf Quasi-Topological Black Holes}. 
Consider a higher-curvature theory of gravity built from arbitrary contractions of the Riemann tensor and the metric and with Lagrangian density $\mathcal{L}(g^{ab},R_{cdef})$.  Quasi-topological gravities are a particular class of this general set of theories that are particularly well-suited for the study of black hole solutions. In essence, the defining property of Quasi-topological gravities is that their equations of motion become of second order when evaluated on static and spherically symmetric (SSS) metrics. This in turn implies uniqueness of SSS solutions and in many cases also a Birkhoff theorem \cite{Oliva:2010eb, Oliva:2011xu, Cisterna:2017umf}, by which those are the only spherically symmetric solutions. 
Lovelock gravities \cite{Lovelock1, Lovelock2} are naturally included in this family of theories, but unlike the former, Quasi-topological gravities exist at all orders in the curvature in any dimension $D\ge 5$~\cite{Bueno:2019ycr, Bueno:2022res, Moreno:2023rfl}. In fact, the family of Quasi-topological gravities is broad enough that it provides a basis of polynomial densities for a effective-field-theory expansion of GR \cite{Bueno:2019ltp}, as we discuss in more detail in the Supplemental Material.
Thus, these theories allow us to capture general EFT corrections to GR, and at the same time they allow us to study black hole solutions in a very robust way. Therefore, they provide the most reliable framework to understand the effects of higher-curvature corrections near black hole singularities.


Now, at every order in the curvature $n$, several Quasi-topological densities exist. However, when restricted to a SSS metric, all of these densities contribute in the same way to the equations of motion. Hence, to study spherically symmetric black hole solutions, it suffices to include one (non-trivial) Quasi-topological density in the action at every order. Let us call $\mathcal{Z}_{n}$ this representative order-$n$ density. For instance, the quadratic density is the Gauss-Bonnet invariant, $\mathcal{Z}_{2}\propto \mathcal{X}_{4}$, the cubic density $\mathcal{Z}_{3}$ can be chosen as the Lagrangian originally found in \cite{Oliva:2010eb,Quasi,Dehghani:2011vu,Ahmed:2017jod,Cisterna:2017umf}, and explicit expressions of $\mathcal{Z}_{n}$ for general $n$ were recently found in \cite{Bueno:2019ycr, Bueno:2022res, Moreno:2023rfl}. We provide a few more details on the form of these Lagrangians in the Supplementary Material. 

Let us then write the action of a Quasi-topological theory as
\begin{equation}\label{QTaction}
I_{\rm QT}=\frac{1}{16\pi G} \int \mathrm{d}^Dx \sqrt{|g|} \left[R+\sum_{n=2}^{n_{\rm max}} \alpha_n \mathcal{Z}_n \right]\, , 
\end{equation}
where  $\alpha_n$ are arbitrary coupling constants with dimensions of length$^{2(n-1)}$. We consider a general SSS ansatz, which can be written as
\be  \label{ff}
\mathrm{d} s^2 = - N(r)^2f(r) \mathrm{d} t^2 + \frac{\mathrm{d} r^2}{f(r)} + r^2 \mathrm{d}  \Omega^2_{D-2} \, ,
\ee
with two undetermined functions $N(r)$ and $f(r)$. The equations of motion of \req{QTaction} imply that 
\begin{equation}
\frac{dN}{dr}=0\, ,\quad \frac{d}{dr}\left[r^{D-1}h(\psi)\right]=0\, ,
\end{equation}
where 
\be \label{eom_psi}
h(\psi) \equiv \psi + \sum_{n=2}^{n_{\rm max}} \alpha_n \psi^n\, ,\quad \psi \equiv  \, \frac{1-f(r)}{r^2}  \, .
\ee
Therefore, the solution has $N(r)=1$ (required by normalization of the time coordinate at infinity) while $f(r)$ is determined by the algebraic equation
\be \label{eom}
h(\psi)  = \frac{m}{r^{D-1}}  \, ,
\ee
where $m$ is an integration constant which is proportional to the ADM mass of the solution. As previously anticipated, this is the only SSS solution of the theory.  Naturally, when $\alpha_n=0$ we recover the Schwarzschild solution, $f=1-m/r^{D-3}$, but let us analyze the effects of the corrections.

%

Consider first the case where a finite number of terms are included, up to $n_{\rm max} = N$. In this case, in the deep black hole interior the highest order contribution dominates and the solution in the vicinity of $r = 0$ is given approximately as
\be 
f = 1 - \left(\frac{m}{\alpha_N} \right)^{1/N} r^{2  - (D-1)/N} + \cdots \, .
\ee
For $N$ finite there is a curvature singularity at $r = 0$ where the Kretschmann scalar diverges. However, something interesting clearly happens as $N$ becomes large and tends toward infinity. Namely, the metric approaches that of an (anti) de Sitter universe with cosmological length scale set by the ratio of the mass parameter and the coupling constant. In the strict limit when an infinite number of higher-curvature terms are included, the interior becomes exactly (anti) de Sitter and the singularity is resolved. 

Note that this fact relies crucially on the ability to take $N$ arbitrarily larger than $D$. For example, for Lovelock theory of order $N$ one must necessarily have $D \ge 2N+1$. In such a case, resolution of the singularity would not be possible by the same mechanism since $(D-1)/N$ would never tend to zero.\footnote{The recently constructed lower-dimensional limits of Lovelock theory would in principle avoid this issue allowing the construction to be extended to four-dimensions. However, in the careful regularizations of those theories~\cite{Lu:2020iav, Fernandes:2020nbq, Hennigar:2020lsl}, the four-dimensional limits of the higher-order Lovelock densities do not produce metrics that coincide with the naive regularizations~\cite{Glavan:2019inb} --- see, {\it e.g.},~\cite{Alkac:2022fuc} for more details. This means, in particular, that the results of~\cite{Gao:2020vhw,Colleaux:2020wfv} produce metrics that do not solve the equations of motion of any theory.}

 \begingroup
\setlength{\tabcolsep}{2.9pt} 
\renewcommand{\arraystretch}{3} 
\begin{table*}[t!]
	\centering
	\begin{tabular}{|c|c|c|c|}
	\hline
 $\alpha_{n}$ & $ h(\psi)$ & $f(r)$ & Differentiablility  at $r=0$\\
 \hline \hline
 $\alpha^{n-1}$ & $\displaystyle \frac{\psi}{1-\alpha \psi}$ & $\displaystyle 1 - \frac{m r^2}{r^{D-1} + \alpha m}$ & $\mathcal{C}^{\infty}$ if $D$ odd, $\mathcal{C}^{D+1}$ if $D$ even  \\ \hline
  $\displaystyle\frac{\alpha^{n-1}}{n}$ & $\displaystyle -\frac{\log(1-\alpha\psi)}{\alpha}$ & $\displaystyle1- \frac{r^2}{\alpha } \left(1-e^{-\alpha m/r^{D-1}}\right)$ & $\mathcal{C}^{\infty}$ \\ \hline
    $\displaystyle n\alpha^{n-1}$ & $\displaystyle \frac{\psi}{(1-\alpha \psi)^2}$ & \scriptsize  $\displaystyle 1-\frac{2mr^2}{r^{D-1}+2\alpha m+\sqrt{r^{2 (D-1)}+4 \alpha  m r^{D-1}}}$ & $\mathcal{C}^{\infty}$ if $D=1\operatorname{mod}4$, else $\mathcal{C}^{\lfloor (D+3)/2\rfloor}$ \\  \hline
    $\displaystyle \frac{(1-(-1)^n)}{2}\alpha^{n-1}$ & $\displaystyle \frac{\psi}{1-\alpha^2\psi^2}$ & $\displaystyle 1-\frac{2mr^2}{r^{D-1}+\sqrt{r^{2 (D-1)}+4 \alpha ^2 m^2}}$ & $\mathcal{C}^{\infty}$ if $D$ odd, $\mathcal{C}^{D+1}$ if $D$ even\\ \hline
    $\displaystyle \frac{(1-(-1)^n)\Gamma \left(\frac{n}{2}\right)}{2\sqrt{\pi } \Gamma \left(\frac{n+1}{2}\right)}\alpha^{n-1}$ & $\displaystyle \frac{\psi}{\sqrt{1-\alpha^2\psi^2}}$ & $\displaystyle 1-\frac{mr^2}{\sqrt{r^{2 (D-1)}+\alpha ^2 m^2}}$ & $\mathcal{C}^{\infty}$ \\ \hline
	\end{tabular}
	\caption{We present various explicit examples of regular black hole solutions for different choices of the Quasi-topological couplings $\alpha_n$. The second and third columns correspond, respectively, to the resulting form of the characteristic function $h(\psi)$ and the metric function $f(r)$ when an infinite tower of terms is included. The fourth column describes the differentiability class of the solution encountered at $r=0$. Observe that $\mathcal{C}^{2}$ is sufficient for the spacetime to be regular.}
	\label{table}
\end{table*}
\endgroup

{\bf Conditions for Singularity Resolution}. We now allow for an infinite tower of higher curvature corrections, letting the sum in~\eqref{eom_psi} extend to $n_{\rm max} \to \infty$. Since the coupling constants appearing in the equations of motion are \textit{a priori} completely arbitrary, here we analyse under what conditions black holes with regular interiors exist as solutions to the full theory.

In general, we need to be able to invert \eqref{eom} for all values of $r$. We can guarantee that the inverse exists and is smooth if $h(\psi)$ is a monotonic function and its image includes all positive numbers. Otherwise, singularities will typically arise at finite values of $r$. On the other hand, we need to ensure that the behaviour near $r=0$ is precisely $f=1+\mathcal{O}(r^2)$, as required to have a regular core. This depends on the asymptotic form of the $\alpha_n$ couplings when $n\rightarrow\infty$.  The following conditions are \textit{sufficient} to guarantee regularity of the solution
\begin{equation}\label{regconditions}
\alpha_{n}\ge 0\,\, \forall\, n\,,\quad \lim_{n\rightarrow\infty} (\alpha_{n})^{\frac{1}{n}}=C>0\, .
\end{equation}
The positivity of the couplings naturally implies that the function $h(\psi)$ is monotonic for $\psi>0$ and has an inverse. On the other hand, $\lim_{n\rightarrow\infty}  (\alpha_{n})^{\frac{1}{n}}=C$ implies that the series in \req{eom_psi} has a radius of convergence $\psi_0=1/C$. Since the coefficients are positive, this means that the function $h(\psi)$ has an actual divergence at $\psi=\psi_0$. Therefore, when $r\rightarrow 0$ we have $\psi\rightarrow\psi_0$ and consequently $f\sim 1-\psi_0 r^2$. This guarantees the solution is at least of class $\mathcal{C}^{2}$ at $r=0$ and hence it has finite Riemann curvature. Physically, this is all we need in order for the spacetime to be regular. We note that the solutions are often more smooth than $\mathcal{C}^{2}$, and in many cases they are $\mathcal{C}^{\infty}$, but this depends on the form of the subleading terms in the small $r$ expansion.

A counterexample of this situation happens if the radius of convergence is infinite, so that $C=0$ in \eqref{regconditions}. In that case, $h(\psi)$ diverges only for $\psi\rightarrow\infty$ and hence $\psi$ will have some running with $r$ when $r\rightarrow 0$,  preventing the formation of a perfectly regular core. The most representative example of this is $\alpha_{n}=\alpha^{n-1}/(n-1)!$, so that we get the exponential function $h(\psi)=\psi e^{\alpha\psi}$. In this theory the behaviour near $r=0$ is $f\sim 1+\frac{r^2(D-1)}{\alpha}\log(r)$. The metric is in this case only $\mathcal{C}^{1}$ which leads to a (very mild) time-like singularity at $r=0$. 
 
On the other hand, the conditions \eqref{regconditions} are not \textit{necessary} --- there are even more choices that lead to regular black holes --- they are however very broad and natural conditions. We illustrate in Table~\ref{table} some examples of re-summed theories satisfying these conditions, and now comment on a particular example in more detail.

\begin{figure}
\centering \hspace{0cm}
\includegraphics[width=0.482\textwidth]{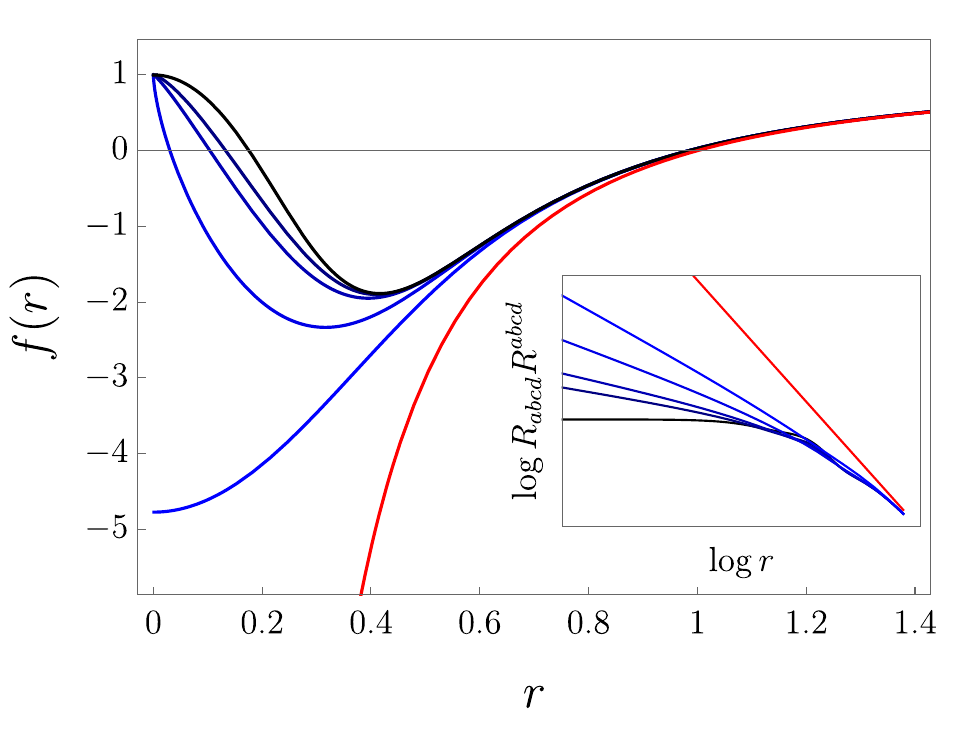}
\caption{We plot the metric function $f(r)$ for various Quasi-topological black holes with $\alpha_n=\alpha^{n-1}$ and $\alpha=0.03$, $m=1$ in $D=5$. The red line represents the Schwarzschild black hole. The increasingly darker blue curves correspond to the solutions with $n_{\rm max}=2,3,5,7$, and the regular Hayward-like black hole, which includes the infinite tower of higher-curvature corrections, is shown in black. The inset plot represents (in a log-log scale) the Kretschmann invariant of the solutions. All solutions have a curvature singularity at the origin with the exception of the Hayward-like one, which tends to a constant value there.  }
\label{fig:hay}
\end{figure}

{\bf Example: Hayward Black Hole}. The simplest re-summation is to take $\alpha_n = \alpha^{n-1}$. The field equation then reduces to the geometric series, which can be summed to give
\be 
 \frac{\psi}{1-\alpha \psi} = \frac{m}{r^{D-1}} \, .
\ee
The metric function is then simply
\be 
f = 1 - \frac{m r^2}{r^{D-1} + \alpha m} \, ,
\ee
which we plot in Fig.\, \ref{fig:hay} in the $D=5$ case.
The black hole is manifestly regular, with the metric $C^\infty$ in the neighbourhood of $r = 0$ if $D$ is odd, and $C^{D+1}$ if $D$ is even. In fact, this is the natural higher-dimensional generalization of the Hayward black hole --- indeed, substituting $D = 4$ in the above gives precisely Hayward's metric~\cite{Hayward:2005gi}. Here we have obtained this metric for any dimension $D \ge 5$ as the solution of a purely gravitational theory that includes an infinite tower of higher-curvature corrections.  To the best of our knowledge, this represents the first time the Hayward black hole has been obtained as a solution to a gravitational theory without additional matter degrees of freedom.



{\bf Regular Black Hole Thermodynamics}. In all cases, it is possible to compute the thermodynamic parameters of the regular black holes in a closed and universal form. This is made possible with the results of~\cite{Bueno:2019ycr,Bueno:2022res}, where the thermodynamics of Quasi-topological black holes were studied for arbitrary functions $h(\psi)$. The thermodynamic potentials read:
\begin{align}
M &= \frac{(D-2) \Omega_{D-2} r_+^{D-1}}{16 \pi G} h(\psi_+) \, ,
\\
T &= \frac{1}{4 \pi r_+ }\left[ \frac{(D-1) r_+^2 h(\psi_+)}{h'(\psi_+)} - 2 \right] \, ,
\\
S &= -\frac{(D-2) \Omega_{D-2}}{8 G} \int \frac{h'(\psi_+)}{\psi_+^{D/2}} {\rm d} \psi_+ \, .
\end{align}
Here $M$ is the ADM mass, $T$ is the Hawking temperature computed via the ordinary Euclidean trick, and $S$ is the Wald entropy. We have introduced the short-hand $\psi_+$, which is just the evaluation of~\eqref{eom_psi} at the horizon, and $\Omega_{D-2} =2 \pi^{(D-1)/2} / \Gamma\left[(D-1)/2\right]$ is the volume of the $(D-2)$ sphere.

It is straightforward to verify that these expressions satisfy the first law of the thermodynamics 
\be 
\mathrm{d}M = T \mathrm{d} S
\ee
irrespective of the choice of function $h(x)$. As an example, the thermodynamic parameters of the Hayward black hole in $D = 5$ read:
\begin{align}
G M &= \frac{3 \pi r_+^4}{8 \left(r_+^2 - \alpha\right)} \, ,
\quad T= \frac{r_+^2 - 2 \alpha}{2 \pi r_+^3} \, ,
\\
G S &= \frac{\pi^2 r_+^3}{2} + 3 \pi^2 \alpha r_+ - \frac{3 \pi^2 \alpha^2 r_+}{4(r_+^2 - \alpha)} 
\nonumber\\
&- \frac{15 \pi^2 \alpha^{3/2}}{4} {\rm arctanh} \left(\frac{\sqrt{\alpha}}{r_+} \right) \, .
\end{align}
It can be easily checked that these quantities satisfy the first law and reduce to the standard Einstein gravity results when $r_+ \gg \sqrt{\alpha}$. Moreover, the mass and entropy are manifestly positive in all cases for which there is a horizon.

It is worth emphasizing here that in this set up, the thermodynamics of the regular black holes is completely unambiguous, arising as the ordinary thermodynamics of a black hole in a higher-curvature theory of gravity. There are many examples in the literature where the study of regular black holes is argued to result in modified thermodynamic relations, or no explanation is given for why the entropy differs from the standard area law. However, it seems likely that those difficulties arise because the thermodynamics studied are \textit{constrained}, \textit{i.e.} the theories studied do not admit a first law of full co-homogeneity due to certain fixed relationships that hold between parameters of the solution --- see, \textit{e.g.},~\cite{Simovic:2023yuv}.

{\bf Discussion}. We have shown for the first time that the Schwarzschild singularity is generically resolved by a purely gravitational mechanism, owing to the inclusion of an infinite tower of higher-curvature terms in the action. The singularity resolution depends crucially on the inclusion of an \textit{infinite} number of such terms, as it remains in any finite truncation of the theory --- although it becomes increasingly weaker with the order of the truncation. An infinite tower of higher-derivative terms is a low energy prediction common to many approaches to quantum gravity, and our construction reveals how the resummation of these terms may eliminate singularities in practice. This is even more relevant since the theories we have studied, while chosen for their particularly desirable properties, are sufficiently general to provide a basis for (vacuum) gravitational effective field theory in $D \ge 5$. In this sense, our construction may not simply be \textit{a} mechanism by which the Schwarzschild singularity is resolved, it may be \textit{the} mechanism.

There are a number of future directions that follow naturally from our work. First, in the cases we have presented, the black holes have inner horizons which may suffer from mass inflation or quantum instabilities~\cite{Hollands:2019whz}. This is not something particular to our construction, but it has been argued to be a generic feature of regular black hole interiors \cite{Carballo-Rubio:2018pmi,Carballo-Rubio:2021bpr}. However, this may depend on the modified dynamics of the underlying theory, and in this sense, our construction provides a perfect framework to test these possible stability issues. In particular, since our theories satisfy a Birkhoff theorem, studying the stability of the solutions against the collapse of spherical shells of matter should be an accessible problem. Moreover, there may exist multi-parametric resummations of the theory for which the surface gravity of the inner horizon manifestly vanishes, which would avoid mass inflation issues all together \cite{Carballo-Rubio:2022kad}.

Our construction is one for which the thermodynamics of regular black holes is completely unambiguous and under control. It therefore provides a natural setting to explore ideas connecting thermodynamics, phase transitions, and singularity resolution. For simplicity, we have here focused on vacuum solutions that are asymptotically flat. Our construction immediately generalizes to other asymptotics and allows for the inclusion of matter fields. The situation maintains its simplicity if the matter stress tensor satisfies $T_{t}^t = T_{r}^r$, since then the only modification is additional terms on the right-hand side of~\eqref{eom}. For an asymptotically AdS generalization, our model provides a concrete setting where the resolution of black hole singularities can be probed from the perspective of holography. It would also be interesting to understand whether there exists a connection between our resummed theories and non-local gravities. Studies of singularities in non-local theories are often limited to the linearized regime  --- see~\cite{Kolar:2023gqi} for an exception. 

Finally, it would be particularly interesting to extend our construction to four-dimensional space-time. Quasi-topological theories do not exist in $D=4$, which means we no longer have access to theories with analytical black hole solutions. This certainly adds a layer of complication in the study of black holes, but even so we expect the moral of the story --- that higher derivatives can resolve black hole singularities --- to apply as well. Yet, it may be possible to embed our solutions in four dimensions by taking the $D\rightarrow 4$ limit of the corresponding Quasi-topological action analogously to the $D\rightarrow 4$ limit of Gauss-Bonnet gravity \cite{Glavan:2019inb,Lu:2020iav, Fernandes:2020nbq, Hennigar:2020lsl}. This would allow one to obtain the metrics in Table \ref{table} with $D=4$ as exact solutions of certain scalar-tensor theories.  It would be interesting to understand if this limit can be defined in a rigorous way.

\vspace{0.1cm}
\begin{acknowledgments} 
We would like to thank Roberto Emparan, Pedro G.~S.~Fernandes, David Kubiz\v{n}\'{a}k and Robert Mann for useful conversations and comments on this manuscript. RAH  also thanks David Kubiz\v{n}\'{a}k and Marek Li\v{s}ka for discussions on related topics. 
PB was supported by a Ram\'on y Cajal fellowship (RYC2020-028756-I) from Spain's Ministry of Science and Innovation. 
The work of PAC received the support of a fellowship from “la Caixa” Foundation (ID 100010434) with code LCF/BQ/PI23/11970032. 
The work of RAH received the support of a fellowship from ``la Caixa” Foundation (ID 100010434) and from the European Union’s Horizon 2020 research and innovation programme under the Marie Skłodowska-Curie grant agreement No 847648 under fellowship code LCF/BQ/PI21/11830027.

\end{acknowledgments}

\bibliographystyle{JHEP-2}
\bibliography{Gravities}
\noindent 

\appendix
\onecolumngrid \vspace{1.5cm}

\section{Further Details on Quasi-Topological Gravities}

\subsection{Definition of Theories; Equations of Motion}
Consider a higher-curvature theory of gravity built from arbitrary contractions of the Riemann tensor and the metric and with Lagrangian density $\mathcal{L}(g^{ab},R_{cdef})$. In the absence of matter, its equations of motion read
\begin{equation}
P_a^{cde}R_{bcde}-\frac{1}{2}g_{ab}\mathcal{L}-2\nabla^c\nabla^d P_{acdb}=0\, ,  
\end{equation}
where $P^{abcd}\equiv \partial \mathcal{L}/\partial R_{abcd}$. Quasi-topological gravities are a particular class of this general set of theories characterized by having 
\be 
\nabla^d P_{acdb}=0
\ee
for general static, spherically (planar, or hyperbolic) symmetric metrics. Namely, when restricted to spherical symmetry, the third term in the equations of motion is absent.

The simplest way to define Quasi-topological gravities at all orders is via a recursion relation, as first developed in~\cite{Bueno:2019ycr}. The recursive formula reads
\begin{align}\label{recursive} 
\tilde{\mathcal{Z}}_{n+5} &= -\frac{3 (n+3)}{2 (n+1) D (D-1)} \tilde{\mathcal{Z}}_{1} \tilde{\mathcal{Z}}_{n+4} + \frac{3 (n+4)}{2 n D (D-1)} \tilde{\mathcal{Z}}_{2} \tilde{\mathcal{Z}}_{n+3} 
- \frac{(n+3)(n+4)}{2n (n+1)D(D-1)} \tilde{\mathcal{Z}}_{3} \tilde{\mathcal{Z}}_{n+2}  \, .  
\end{align}
Given five non-trivial densities, $\{\tilde{\mathcal{Z}}_1,\,\cdots , \tilde{\mathcal{Z}}_{5}\}$ this recursive relation can be used to generate Quasi-topological theories at all orders. Note that here we have used $\tilde{\mathcal{Z}}_n$ to indicate that we are matching the conventions of~\cite{Bueno:2019ycr} in this expression. The $\tilde{\mathcal{Z}}_n$ densities are related to those we study here by a simple rescaling:
\be 
\mathcal{Z}_n = -\frac{1}{(D-2n)} \tilde{\mathcal{Z}}_n \, .
\ee
Note that the $\tilde{\mathcal{Z}}_n$ density in $D=2n$ dimensions does not contribute to the spherically symmetric equations of motion (hence the name, \textit{Quasi-topological}), as signaled by the divergence of the normalization factor $1/(D-2n)$. Thus, in even dimension $D=2n$ one should remove the $n$-th term, $\alpha_{n}$, from the $h(\psi)$ function defined in the main text. This would complicate the form of some of the explicit solutions we give in even dimensions, but it does not affect our conclusions so we choose not to deliberate on this issue.


An interesting observation we have made in the course of preparing this manuscript is that~\eqref{recursive} appears to imply a Birkhoff theorem at all orders, provided the five `seed' theories also satisfy a Birkhoff theorem. We have established this in $D = 5$ and $D = 6$ and it appears the general mechanism will extend to all dimensions $D \ge 5$ (though explicitly proving this would be an important development). 

The existence of five Quasi-topological theories in each dimension $D \ge 5$ was proven in~\cite{Bueno:2019ycr} --- though see also~\cite{Moreno:2023rfl} for recent work on this. A particular basis of five such densities reads:
\begin{align}
\tilde{\mathcal{Z}}_1 &= -R \, ,
\\
\tilde{\mathcal{Z}}_2 &= -\frac{1}{(D-2)(D-3)} \left[R^2 - 4 R_{ab} R^{ab} + R_{abcd}R^{abcd} \right]
\\
\tilde{\mathcal{Z}}_3 &= - \frac{8 (2D-3)}{(D-2)(D-3)(D - 4)(3D^2 - 15D + 16)} \Bigg[ (D - 4){{{R_a}^b}_c{}^d} {{{R_b}^e}_d}^f {{{R_e}^a}_f}^c  
               +  \frac{3(3D - 8)}{8(2D - 3)} R_{a b c d} R^{a b c d} R  
               \nonumber\\
               &- \frac{3(3D-4)}{2(2D - 3)} {R_a}^c {R_c}^a R  - \frac{3(D-2)}{(2D - 3)} R_{a c b d} {R^{a c b}}_e R^{d e} + \frac{3D}{(2D - 3)} R_{a c b d} R^{a b} R^{c d} \nonumber\\
              & 
                + \frac{6(D-2)}{(2D - 3)} {R_a}^c {R_c}^b {R_b}^a  + \frac{3D}{8(2D - 3)} R^3  \Bigg]\, ,
\\
\tilde{\mathcal{Z}}_4 &=  -\frac{384 (D-8) R^a_b R_a^c R_c^d R^b_d}{(D-2)^5(D^3 - 8 D^2 + 48 D - 96)} - \frac{1152 R_{ab}R^{ab}R_{cd}R^{cd}}{(D-2)^5(D^3 - 8 D^2 + 48 D - 96)} 
\nonumber\\
&- \frac{64(D^3 - 10D^2 + 40D + 24) R R_a^c R_c^b R_b^a }{(D-1)(D-2)^5(D^3 - 8 D^2 + 48 D - 96)} + \frac{24(D^4 - 6 D^3 + 20 D^2 + 104 D - 64) R^2 R_{ab}R^{ab}}{(D-1)^2(D-2)^5(D^3 - 8 D^2 + 48 D - 96)} 
\nonumber\\
&- \frac{(D^5 + 6 D^4 - 64 D^3 + 416 D^2 + 176 D - 480) R^4}{(D-1)^3(D-2)^5(D^3 - 8 D^2 + 48 D - 96)} - \frac{96(D+2) R R^{ab}R^{cd} W_{acbd}}{(D-1)(D-2)^4(D-3)(D-4)} 
\nonumber\\
&- \frac{6(2D^5 - D^4 - 31 D^3 + 20 D^2 + 20 D - 16) R^2 W_{abcd}W^{abcd}}{(D-1)^2(D-2)^3(D-3)(D-4)(2D^4 - 17 D^3 + 49 D^2 - 48 D + 16)} 
\nonumber\\
&+ \frac{96(2 D^4 - 7 D^3 - 7 D^2 + 18 D - 8) R R_b^a W_{ac}{}^{de} W_{de}{}^{bc}}{(D-1)(D-2)^3(D-3)(D-4)(2 D^4 - 17 D^3 + 49 D^2 - 48 D + 16)} 
\nonumber\\
&+ \frac{384 R_a^c R^{ab} R^{de} W_{bdce}}{(D-2)^4(D-3)(D-4)} - \frac{48 (7D^2 - 10D + 4) R^{ab}R^{cd}W_{ac}{}^{ef} W_{bd ef}}{(D-2)^3(D-3)(2 D^4 - 17 D^3 + 49 D^2 - 48 D + 16)} 
\nonumber\\
&- \frac{8 (2 D^4 - 15 D^3 + 26 D^2 + 27 D - 58) R W_{ab}{}^{ef} W^{abcd} W_{cdef}}{(D-1)(D-2)^2(D-3)(D-4) (D^2 - 6 D + 11)(D^3 - 9 D^2 + 26 D - 22)} 
\nonumber\\
&- \frac{48 (7 D^2 - 10 D + 4) R_a^c R^{ab} W_b{}^{def}W_{cdef}}{(D-2)^3(D-3)(2 D^4 - 17 D^3 + 49 D^2 - 48 D + 16)} + \frac{96 R^{ab} W_a{}^{cde}W_{bc}{}^{fg}W_{defg}}{(D-2)^2(D-3)(D-4)(D^2 - 6 D + 11)} 
\nonumber\\
&- \frac{3(3D-4) W_{ab}{}^{cd} W_{cd}{}^{ef} W_{ef}{}^{gh} W_{gh}{}^{ab}}{(D-2)(D-3) (D^5 - 14 D^4 + 79 D^3 -224 D^2 + 316 D - 170)}\, ,               
\end{align}

{\tiny
\begin{align}
\tilde{\mathcal{Z}}_5 &= +\frac{512 (-64 - 12 D + D^2) R_{a}{}^{c} R^{ab} R_{b}{}^{d} R_{c}{}^{e} R_{de}}{(-4 + D) (-2 + D)^6 (-128 + 32 D + D^3)} + \frac{5120 (4 + D) R_{ab} R^{ab} R_{c}{}^{e} R^{cd} R_{de}}{(-4 + D) (-2 + D)^6 (-128 + 32 D + D^3)} 
\nonumber\\
&-  \frac{640 (4608 + 3712 D - 2880 D^2 + 664 D^3 - 126 D^4 + 7 D^5) R_{a}{}^{c} R^{ab} R_{b}{}^{d} R_{cd} R}{(-4 + D) (-2 + D)^6 (-1 + D) (-128 + 32 D + D^3) (-96 + 48 D - 8 D^2 + D^3)}
\nonumber\\
& -  \frac{1920 (-768 - 320 D + 280 D^2 - 58 D^3 + 11 D^4) R_{ab} R^{ab} R_{cd} R^{cd} R}{(-4 + D) (-2 + D)^6 (-1 + D) (-128 + 32 D + D^3) (-96 + 48 D - 8 D^2 + D^3)} 
\nonumber\\
&-  \frac{160 (4096 - 27136 D + 7168 D^2 + 160 D^3 - 64 D^4 + 116 D^5 - 16 D^6 + D^7) R_{a}{}^{c} R^{ab} R_{bc} R^2}{(-4 + D) (-2 + D)^6 (-1 + D)^2 (-128 + 32 D + D^3) (-96 + 48 D - 8 D^2 + D^3)} 
\nonumber\\
&+ \frac{40 (30720 - 20992 D - 41216 D^2 + 17920 D^3 - 2784 D^4 + 656 D^5 + 28 D^6 - 8 D^7 + D^8) R_{ab} R^{ab} R^3}{(-4 + D) (-2 + D)^6 (-1 + D)^3 (-128 + 32 D + D^3) (-96 + 48 D - 8 D^2 + D^3)} 
\nonumber\\
&-  \frac{(155648 + 231424 D - 530176 D^2 + 136384 D^3 - 14336 D^4 + 4272 D^5 + 1296 D^6 - 204 D^7 + 16 D^8 + D^9) R^5}{(-4 + D) (-2 + D)^6 (-1 + D)^4 (-128 + 32 D + D^3) (-96 + 48 D - 8 D^2 + D^3)}
\nonumber\\
& -  \frac{240 (-80 - 100 D + 8 D^2 + 7 D^3) R^{ab} R^{cd} R^2 W_{acbd}}{(-4 + D) (-3 + D) (-2 + D)^5 (-1 + D)^2 (96 - 48 D + 7 D^2)} 
\nonumber\\
&-  \frac{10 (-128 + 896 D - 2552 D^2 + 3900 D^3 - 2970 D^4 + 425 D^5 + 710 D^6 - 243 D^7 - 10 D^8 + 8 D^9) R^3 W_{abcd} W^{abcd}}{(-4 + D) (-3 + D) (-2 + D)^4 (-1 + D)^3 (16 - 48 D + 49 D^2 - 17 D^3 + 2 D^4) (16 - 56 D + 69 D^2 - 27 D^3 + 4 D^4)} 
\nonumber\\
&+ \frac{240 (64 - 256 D + 260 D^2 + 292 D^3 - 795 D^4 + 516 D^5 - 35 D^6 - 42 D^7 + 8 D^8) R^{ab} R^2 W_{a}{}^{cde} W_{bcde}}{(-4 + D) (-3 + D) (-2 + D)^4 (-1 + D)^2 (16 - 48 D + 49 D^2 - 17 D^3 + 2 D^4) (16 - 56 D + 69 D^2 - 27 D^3 + 4 D^4)} 
\nonumber\\
&+ \frac{1920 (-48 - 14 D + 7 D^2) R_{a}{}^{c} R^{ab} R^{de} R W_{bdce}}{(-4 + D) (-3 + D) (-2 + D)^5 (-1 + D) (96 - 48 D + 7 D^2)} 
\nonumber\\
&-  \frac{240 (128 - 704 D + 1464 D^2 - 1240 D^3 + 60 D^4 + 503 D^5 - 221 D^6 + 28 D^7) R^{ab} R^{cd} R W_{ac}{}^{ef} W_{bdef}}{(-4 + D) (-3 + D) (-2 + D)^4 (-1 + D) (16 - 48 D + 49 D^2 - 17 D^3 + 2 D^4) (16 - 56 D + 69 D^2 - 27 D^3 + 4 D^4)}
\nonumber\\
& -  \frac{11520 R_{a}{}^{c} R^{ab} R_{d}{}^{f} R^{de} W_{becf}}{(-3 + D) (-2 + D)^5 (96 - 48 D + 7 D^2)} 
\nonumber\\
&-  \frac{20 (3632 - 7644 D - 4296 D^2 + 23905 D^3 - 23526 D^4 + 8466 D^5 + 560 D^6 - 1437 D^7 + 478 D^8 - 70 D^9 + 4 D^{10}) R^2 W_{ab}{}^{ef} W^{abcd} W_{cdef}}{(-4 + D) (-3 + D) (-2 + D)^3 (-1 + D)^2 (11 - 6 D + D^2) (-22 + 26 D - 9 D^2 + D^3) (176 - 600 D + 775 D^2 - 482 D^3 + 161 D^4 - 28 D^5 + 2 D^6)}
\nonumber\\
& -  \frac{240 (128 - 704 D + 1464 D^2 - 1240 D^3 + 60 D^4 + 503 D^5 - 221 D^6 + 28 D^7) R_{a}{}^{c} R^{ab} R W_{b}{}^{def} W_{cdef}}{(-4 + D) (-3 + D) (-2 + D)^4 (-1 + D) (16 - 48 D + 49 D^2 - 17 D^3 + 2 D^4) (16 - 56 D + 69 D^2 - 27 D^3 + 4 D^4)} 
\nonumber\\
&-  \frac{15360 R_{a}{}^{c} R^{ab} R_{b}{}^{d} R^{ef} W_{cedf}}{(-3 + D) (-2 + D)^5 (96 - 48 D + 7 D^2)} 
\nonumber\\
&+ \frac{960 (4 - 12 D + 11 D^2) R_{a}{}^{c} R^{ab} R^{de} W_{bd}{}^{fg} W_{cefg}}{(-4 + D) (-3 + D) (-2 + D)^4 (16 - 56 D + 69 D^2 - 27 D^3 + 4 D^4)} 
\nonumber\\
&+ \frac{160 (-232 + 550 D - 253 D^2 - 242 D^3 + 221 D^4 - 62 D^5 + 6 D^6) R^{ab} R W_{a}{}^{cde} W_{bc}{}^{fg} W_{defg}}{(-4 + D) (-3 + D) (-2 + D)^3 (-1 + D) (11 - 6 D + D^2) (176 - 600 D + 775 D^2 - 482 D^3 + 161 D^4 - 28 D^5 + 2 D^6)} 
\nonumber\\
&+ \frac{320 (4 - 12 D + 11 D^2) R_{a}{}^{c} R^{ab} R_{b}{}^{d} W_{c}{}^{efg} W_{defg}}{(-4 + D) (-3 + D) (-2 + D)^4 (16 - 56 D + 69 D^2 - 27 D^3 + 4 D^4)} 
\nonumber\\
&-  \frac{80 (12 - 28 D + 17 D^2) R^{ab} R^{cd} W_{ac}{}^{ef} W_{bd}{}^{gh} W_{efgh}}{(-3 + D) (-2 + D)^3 (176 - 600 D + 775 D^2 - 482 D^3 + 161 D^4 - 28 D^5 + 2 D^6)} 
\nonumber\\
&-  \frac{15 (-528 + 482 D + 241 D^2 - 425 D^3 + 194 D^4 - 39 D^5 + 3 D^6) R W_{ab}{}^{ef} W^{abcd} W_{cd}{}^{gh} W_{efgh}}{(-4 + D) (-3 + D) (-2 + D)^2 (-1 + D) (85 - 99 D + 48 D^2 - 11 D^3 + D^4) (-170 + 316 D - 224 D^2 + 79 D^3 - 14 D^4 + D^5)} 
\nonumber\\
&-  \frac{80 (12 - 28 D + 17 D^2) R_{a}{}^{c} R^{ab} W_{b}{}^{def} W_{cd}{}^{gh} W_{efgh}}{(-3 + D) (-2 + D)^3 (176 - 600 D + 775 D^2 - 482 D^3 + 161 D^4 - 28 D^5 + 2 D^6)} 
\nonumber\\
&+ \frac{240 R^{ab} W_{a}{}^{cde} W_{bc}{}^{fg} W_{de}{}^{hi} W_{fghi}}{(-4 + D) (-3 + D) (-2 + D)^2 (85 - 99 D + 48 D^2 - 11 D^3 + D^4)}
\nonumber\\
& -  \frac{4 (-5 + 4 D) W_{ab}{}^{ef} W^{abcd} W_{cd}{}^{gh} W_{ef}{}^{ij} W_{ghij}}{(-3 + D) (-2 + D) (-1150 + 2954 D - 3202 D^2 + 1934 D^3 - 705 D^4 + 155 D^5 - 19 D^6 + D^7)}\, .
\end{align}
}
Explicit expressions for $n$-th order densities were also obtained in \cite{Bueno:2019ycr}, and more recently \cite{Moreno:2023rfl} provided a simpler set of invariants. 
Note that Quasi-topological invariants are not unique --- there exist multiple distinct theories at each order in curvature that are Quasi-topological gravities. However, all such theories differ by \textit{trivial densities} --- curvature invariants which do not contribute to the equations of motion on static, spherically symmetric background. Therefore, when restricted to spherical symmetry, all Quasi-topological theories contribute to the equations of motion in the same way. In this sense, the above densities correspond to particular representatives at each order, but they are not the \textit{only} such representatives at each order.

While direct computation of the covariant field equations is simple at low orders in curvature, the equations of motion at arbitrary orders are most easily obtained via the principle of symmetric criticality using reduced Lagrangian methods~\cite{Fels:2001rv, Deser:2003up}. Namely, one evaluates the action of the theory on a symmetry reduced ansatz of the form
\be 
{\rm d} s^2 = -N(r)^2 f(r) {\rm d} t^2 + \frac{{\rm d} r^2}{f(r)} + r^2 {\rm d} \Omega_{D-2}^2 \, ,
\ee
and subsequently integrates over the angular directions to give a reduced action depending on the two functional degrees of freedom $(N, f)$. After integrating by parts and discarding total derivatives, the reduced action for all Quasi-topological theories takes the form
\be 
I_{\rm QT}[N,f] = \Omega_{D-2} \int {\rm d}t {\rm d}r N(r) \left[r^{D-1} h(\psi) \right]' \, ,
\ee
where the prime denotes the radial derivative. Variation of the action with respect to $N$ and $f$ yields two components of the field equations $\mathcal{E}_{ab}$,
\be 
\frac{1}{\Omega_{D-2} r^{D-2}} \frac{\delta I[N, f]}{\delta N} = \frac{2 \mathcal{E}_{tt}}{f N^2} \, , \quad \frac{1}{\Omega_{D-2} r^{D-2}} \frac{\delta I[N, f]}{\delta f} = \frac{\mathcal{E}_{tt}}{N f^2} + N \mathcal{E}_{rr} \, .
\ee
The angular components of the field equations are automatically satisfied provided $\mathcal{E}_{tt}$ and $\mathcal{E}_{rr}$ are due to the generalized Bianchi identity $\nabla_a \mathcal{E}^{ab} = 0$~\cite{Bueno:2017sui}. Applying this to the reduced action for Quasi-topological theories quoted above one obtains
\be 
\frac{\delta I[N, f]}{\delta f}=0 \, \, \Rightarrow \, \,  h'(\psi) N'(r) = 0 \, , \quad \frac{\delta I[N, f]}{\delta N}=0 \, \, \Rightarrow \, \,  \left[r^{D-1} h(\psi) \right]' = 0 \, ,
\ee
from which the solutions in the main text follow.

\subsection{Quasi-topological gravities and EFT}
In an EFT extension of GR, one must add all the invariants formed with Weyl curvature to the Einstein-Hilbert action. Operators containing Ricci curvature can be removed by field redefinitions since Ricci curvature vanishes on vacuum solutions of the two-derivative theory.  We focus on the operators that are polynomial in the curvature. At order $n$ in the curvature we have a number $N(n)$ of independent Weyl invariants $\mathcal{L}_{n}^{(i)}\sim (W_{abcd})^{n}$, $i=1,\ldots, N(n)$, where each invariant differs in the way the indices are contracted. Now, on a static and spherically symmetric metric it turns out that all such Weyl invariants become proportional to each other \cite{Deser:2005pc}, 
\begin{equation}
\mathcal{L}_{n}^{(i)}\Big|_{\rm SSS}=a_{i} F(N,f)^n\, ,
\end{equation}
where $F(N,f)$ is a fixed quantity that depends on the functions $N(r)$ and $f(r)$ of the metric and $a_{i}$ is a constant that depends on the particular Weyl invariant. Then, if for instance $a_{1}\neq 0$,  we can consider an alternative basis of invariants given by
\begin{equation}
\tilde{\mathcal{L}}_{n}^{(1)}=\mathcal{L}_{n}^{(1)}\, ,\quad \tilde{\mathcal{L}}_{n}^{(i)}=\mathcal{L}_{n}^{(i)}-\frac{a_i}{a_1}\mathcal{L}_{n}^{(1)}\, , \quad i=2,\ldots, N(n)\, .
\end{equation}
In this basis we have $\tilde{\mathcal{L}}_{n}^{(i)}\big|_{\rm SSS}=0$ for $i\neq 1$, and therefore only the density $\tilde{\mathcal{L}}_{n}^{(1)}$ contributes to the equations of motion of spherically symmetric spacetimes. This proves that there is only one way of modifying the Schwarzschild solution at each order in the curvature. Finally, we can ``complete''  $\tilde{\mathcal{L}}_{n}^{(1)}$ to a Quasi-topological density by introducing terms with Ricci curvature via field redefinitions, as shown in \cite{Bueno:2019ltp}. Therefore, Quasi-topological gravities capture the most general correction to the Schwarzschild solution coming from polynomial operators in the EFT of gravity. 
The status of non-polynomial densities --- \textit{i.e.}, those with covariant derivatives of the Weyl tensor --- is less clear, but the results of \cite{Bueno:2019ltp} show that all densities with two covariant derivatives and an arbitrary number of Weyl tensors can again be mapped to Quasi-topological gravities.


  
\end{document}
%